\documentclass{ws-ijmpa}

\begin{document}

\markboth{P. Lebiedowicz and A. Szczurek}
{Exclusive production of $\pi^{+}\pi^{-}$ pairs
in proton-proton and proton-antiproton collisions}

%
\catchline{}{}{}{}{}
%

\title{EXCLUSIVE PRODUCTION OF $\pi^{+}\pi^{-}$
PAIRS IN PROTON-PROTON AND PROTON-ANTIPROTON COLLISIONS}

\author{PIOTR LEBIEDOWICZ}
\address{Institute of Nuclear Physics PAN, PL-31-342 Cracow, Poland\\
piotr.lebiedowicz@ifj.edu.pl}

\author{ANTONI SZCZUREK}
\address{University of Rzesz\'ow, PL-35-959 Rzesz\'ow, Poland\\
Institute of Nuclear Physics PAN, PL-31-342 Cracow, Poland\\
antoni.szczurek@ifj.edu.pl}

\maketitle


\begin{abstract}
We report on a detailed investigation of four-body
$p p \to p p \pi^+ \pi^-$ and $p \bar{p} \to p \bar{p} \pi^+ \pi^-$) 
reactions which constitute an irreducible background to 
three-body processes $p p \to p p M$,
where $M$ is a broad resonance in the $\pi^+ \pi^-$ channel, e.g. 
$M=\sigma, \rho^{0}, f_{0}(980), f_{2}(1275), f_{0}(1500)$.
We include double-diffractive contribution
(both pomeron and reggeon exchanges) as well as the pion-pion 
rescattering contributions.
The first process dominates at higher energies and small 
pion-pion invariant masses while the second becomes important at 
lower energies and higher pion-pion invariant masses.
We compare our results with the experimental data.
We make predictions for future experiments at PANDA, RHIC, Tevatron and LHC energies.
The two-dimensional distribution in rapidity space of pions $(y_{\pi^+}, y_{\pi^-})$
is particularly interesting.
The higher the incident energy, the higher preference
for the same-hemisphere emission of pions.
The processes considered constitute a sizeable contribution to
the total nucleon-nucleon cross section as well as to pion
inclusive cross section.

\end{abstract}

\ccode{PACS numbers: 11.55.Jy, 13.75.Cs, 13.75.Lb, 13.85.Lg}

\section{Introduction}	

Diffractive processes are very attractive from 
the general point of view of the reaction mechanism.
Recently there is a growing interest in understanding
exclusive three-body reactions $p p \to p p M$ at high energies,
where the resonance $M$ is produced in the central rapidity region. 
These resonances are seen (or will be seen) "on" the background of 
a $\pi\pi$ continuum.
\footnote{In generality, the resonance and continuum contributions 
may interfere and produce even a dip.
A good example is interference $f_{0}(980)$ and $\sigma$ mesons (see \cite{LSK09,Alde97}).}
For example, the two-pion background to
exclusive production of $f_{0}(1500)$ meson,
was already discussed in \cite{SL09}.
At larger energies two-pomeron exchange mechanism
dominates in central production \cite{AG81}.
In calculating the amplitude related to 
double diffractive mechanism for $p p \to p p \pi^+ \pi^-$
we follow the general rules of Pumplin and Henyey \cite{PH76}.

\section{Results and Conclusions}

\begin{figure}[t]
\begin{center}
\parbox[c]{0.45\textwidth}{
\centerline{\psfig{file=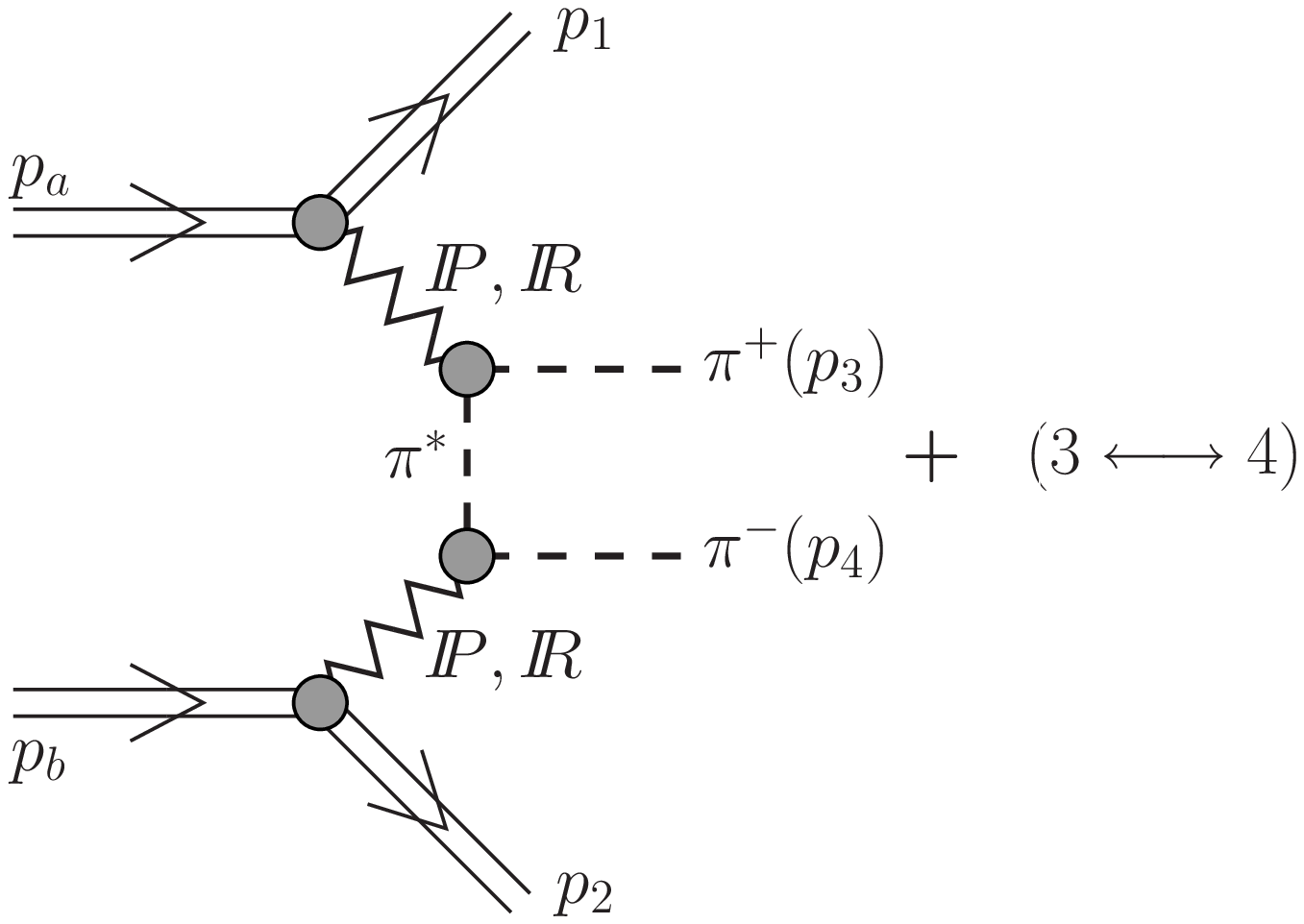,width=5cm}}
}
\parbox[c]{0.45\textwidth}{
\centerline{\psfig{file=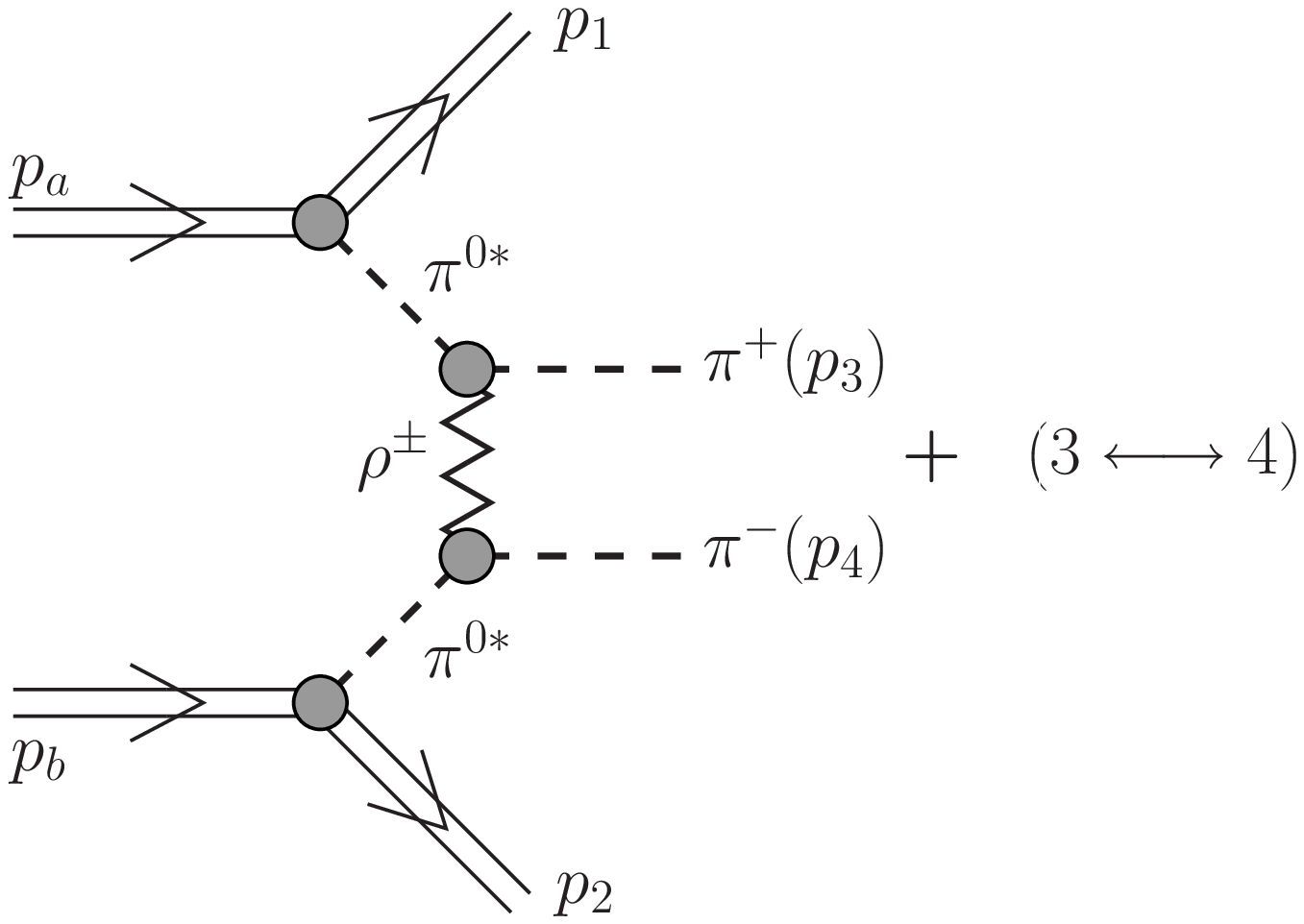,width=5cm}}
}\qquad
\caption{Diagrams representing mechanisms 
for exclusive production of $\pi^{+}\pi^{-}$ pairs at high energies.
{\bf Left}: 
Central double-diffractive mechanism with pomeron and reggeon exchanges.
{\bf Right}: 
Pion-pion rescattering mechanism (here only $\rho$-reggeon exchange is relevant). 
}
\label{f1}
\end{center}
\end{figure}

The results presented here are based on Refs \cite{LS10,LSK09},
where both double diffractive and pion-pion rescattering processes were considered in detail. 
These mechanisms of exclusive production of $\pi^{+}\pi^{-}$ pairs 
at high energies are depicted in Fig.~\ref{f1}.

In the first case the energy dependence of the amplitudes of $\pi N$
subsystems was parametrized in the Regge form with pomeron and 
reggeon exchanges.
The strength parameters and values of the pomeron and reggeon 
trajectories are
taken from the Donnachie-Landshoff analysis \cite{DL92} 
for total cross section for $\pi N$ scattering. 
The slope parameters, are adjusted to the existing experimental data 
for elastic $\pi N$ scattering (see details in \cite{LS10}).
There is a region of energies where the interference term dominates. 
We nicely describe the data for elastic $\pi N$ scattering 
for $\sqrt{s} >$ 2.5 GeV.

In the second case the pion-pion amplitude was parametrized 
using a recent phase shift analysis at the low pion-pion energies \cite{LSK09}
and a Regge form of the continuum obtained by the assumption
of Regge factorization \cite{SNS02}.
The two contributions occupy slightly different parts of the phase space, 
have different energy dependence and in principle can
be resolved experimentally. The interference of amplitudes
of the both processes is almost negligible.

\begin{figure}[t]
\begin{center}
\parbox[c]{0.6\textwidth}{
\centerline{\psfig{file=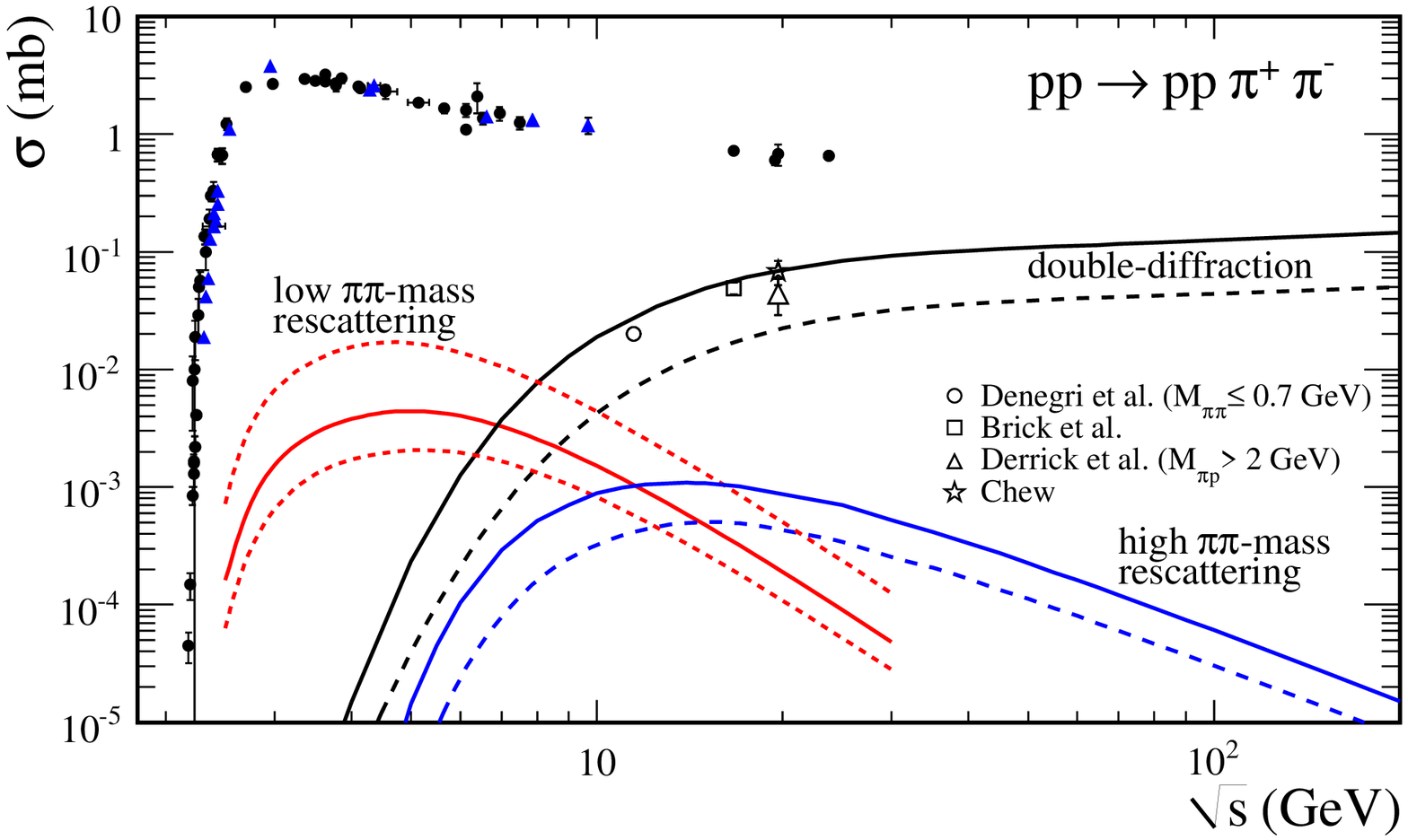,width=7.5cm}}
}
\parbox[c]{0.35\textwidth}{
\centerline{\psfig{file=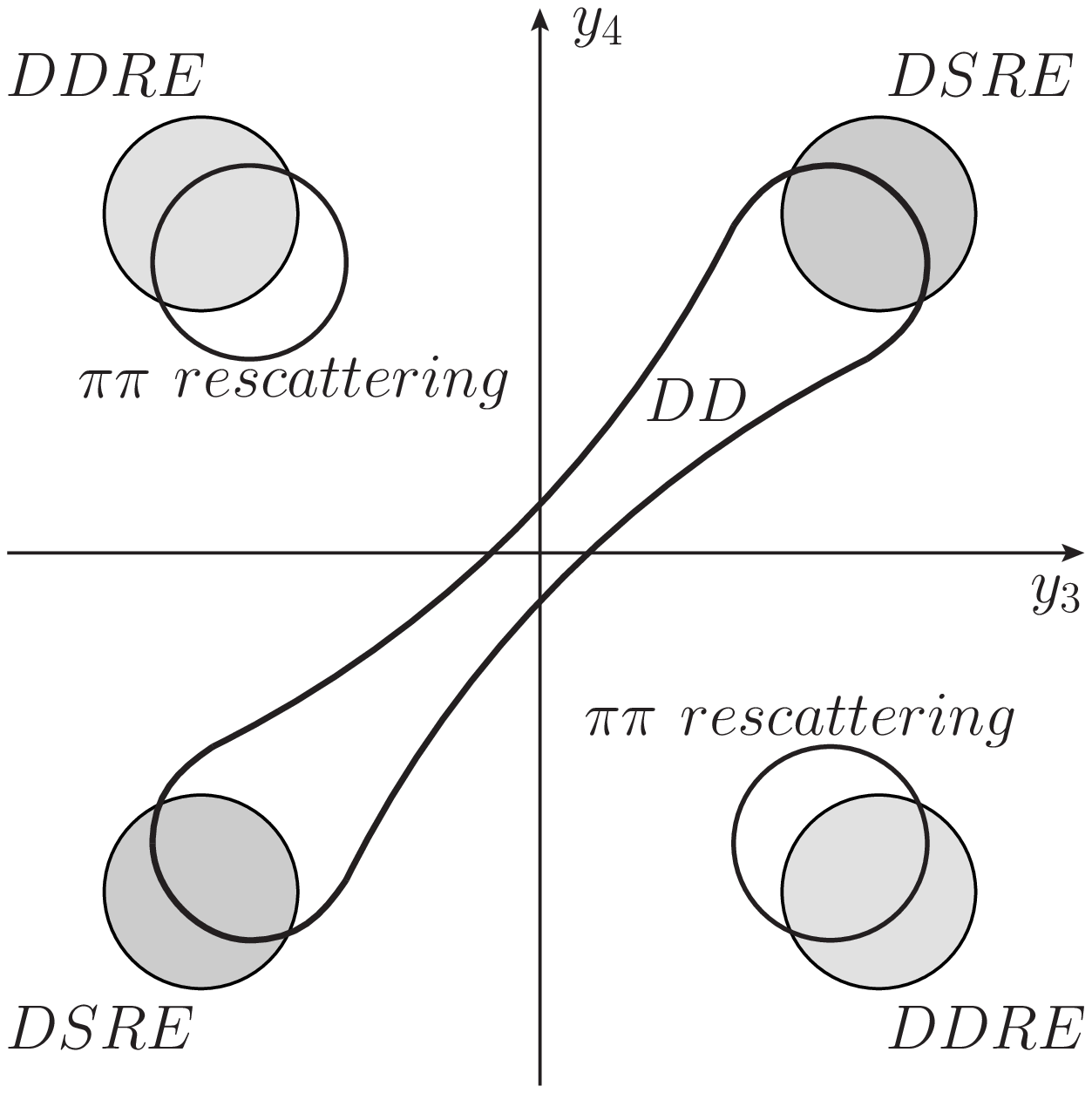,width=4cm}}
}
\caption{
{\bf Left}: 
Cross section for the $p p \to p p \pi^+ \pi^-$ reaction
integrated over phase space as a function of the center-of-mass energy.
We compare the pion-pion rescattering contributions obtained from 
partial wave analysis 
(low $\pi\pi$-mass rescattering) and from the Regge phenomenology 
(high $\pi\pi$-mass rescattering)
as well as double-diffractive contribution 
with the experimental data.
The theoretical uncertainties for these contributions are shown in addition.
{\bf Right}: 
A localization of different mechanisms for  
production of $\pi^{+}\pi^{-}$ pairs
in $pp$ and $p\bar{p}$ collisions at high energies.
}
\label{f2}
\end{center}
\end{figure}

In Fig.~\ref{f2} we compare our results with experimental data
for the $pp \to pp\pi^+\pi^-$ reaction (filled circles) 
which are more than 1 mb for ($2.5<\sqrt{s}<10$) GeV.
The integrated over full phase space cross section of the central
double-diffractive component grows slowly with incident energy.
At lower energies the pion-pion rescattering contributions are important,
however, there the production of single and double resonances 
constitutes the dominant mechanism \cite{LSK09}.
The search for a Double Pomeron Exchange (DPE) contribution
led to an upper limits (open symbols) \cite{open_symbols}.

The central double-diffractive contribution (CDD) lays along 
the diagonal $y_3 = y_4$ and the classical DPE in the center 
$y_3 \approx y_4$.
The $I\!\!P \otimes I\!\!P$ cross section peaks at
midrapidities of pions, while $I\!\!P \otimes I\!\!R$ and 
$I\!\!R \otimes I\!\!P$
at backward and forward pion rapidities, respectively. 
When interfering the three components in the amplitude 
produce (camel-like) enhancements 
of the cross section at forward/backward rapidities.
The diffractive single resonance excitation (DSRE) contribution
is situated at the end points of the CDD contribution.
The diffractive double resonance excitation (DDRE) contribution
is expected at ($y_3 \sim y_{beam}$ and $y_4 \sim y_{target}$)
or ($y_3 \sim y_{target}$ and $y_4 \sim y_{beam}$), i.e.
well separated from the CDD contribution.

The exciting experimental possibility of using the STAR detector 
at RHIC have been presented at this workshop in the plenary
talk \cite{Guryn_MESON10}.
The future physics program of Central Production
will focus on particle production resulting 
from the DPE process.



\end{document}